\documentclass[twocolumn,showpacs,superscriptaddress,preprintnumbers,amsmath,amssymb,epsfig,floatfix,prb]{revtex4-1}
\usepackage{epsfig}
\usepackage{graphicx}
\usepackage{dcolumn}
\usepackage{color}
\usepackage{natbib}  
\usepackage{hyperref}
\usepackage{breakurl}
\hypersetup{colorlinks=true, citecolor=blue, urlcolor=blue, linkcolor=blue}
\usepackage{bm}
\usepackage{tabularx}
\newcolumntype{L}[1]{>{\raggedright\arraybackslash}p{#1}}
\newcolumntype{C}[1]{>{\centering\arraybackslash}p{#1}}
\newcolumntype{R}[1]{>{\raggedleft\arraybackslash}p{#1}}

\makeatletter
\renewcommand*{\@fnsymbol}[1]{\ensuremath{\ifcase#1\or  \dagger\or *\or \else\@ctrerr\fi}}
\makeatother

\begin{document}
\title{Magnetic and electronic phase crossovers in graphene nanoflakes}
\author{Shreemoyee Ganguly}
\affiliation{Thematic Unit of Excellence on Computational Materials Science, S. N. Bose National Centre for Basic Sciences, Kolkata 700098, India}
\author{Mukul Kabir}
\affiliation{Department of Physics, Indian Institute of Science Education and Research, Pune 411008, India}
\affiliation{Centre for Energy Science, Indian Institute of Science Education and Research, Pune 411008, India}
\author{Tanusri Saha-Dasgupta}
\affiliation{Thematic Unit of Excellence on Computational Materials Science, S. N. Bose National Centre for Basic Sciences, Kolkata 700098, India}
\affiliation{Department of Condensed Matter Physics and Materials Science, S. N. Bose National Centre for Basic Sciences, Kolkata 700098, India}
\date{\today}

\begin{abstract}  
Manipulation of intrinsic magnetic and electronic structures of graphene nanoflakes is of technological importance. Here we carry out 
systematic study of the magnetic and electronic phases, and its manipulation in graphene nanoflakes employing first-principles calculation. 
We illustrate the intricate shape and size dependence on the magnetic and electronic properties, and further investigate the effects of 
carrier doping, which could be tuned by gate voltage. A transition from nonmagnetic to magnetic phase is observed at a critical flake 
size for the flakes without sublattice imbalance, which we identify to be originated from the armchair defect at the junctions of two 
sublattices on the edge. Electron, or hole doping simultaneously influences the magnetic and electronic structures, and triggers phase 
changes. Beyond a critical doping, crossover from antiferromagnetic to ferromagnetic phase is observed for the flakes without sublattice 
imbalance, while suppression of magnetism, and a possible crossover from magnetic to nonmagnetic phase is observed for flakes with sublattice
imbalance. Simultaneous to magnetic phase changes, a semiconductor to (half) metal transition is observed, upon
carrier doping. 
\end{abstract}
\maketitle

\section{Introduction}
Plausible applications in spin-based electronics and information processing~\citep{10.1038/nphys547,0034-4885-73-5-056501,0022-3727-47-9-094011, nnano.2014.214,srep07634} have made unconventional magnetism in $sp^{2}$ bonded 2D hexagonal network of C atoms in graphene a topic of recent discussion.
Carbon based materials have unique advantages in this regard due to very weak spin-orbit and hyperfine coupling,~\citep{nphys544, nl072667q, PhysRevB.80.155401} and high spin-wave stiffness.~\citep{0953-8984-18-31-016} Thus, magnetic carbon structures are expected to exhibit higher Curie temperature and spin correlation length, compared to conventional magnets.~\citep{PhysRevLett.100.047209} Further, the spin-transport in graphene could be manipulated easily by external perturbations such as electric fields.~\citep{nature05180}

Magnetism in otherwise nonmagnetic graphene can be induced by the presence of point/extended defects~\citep{PhysRevB.75.125408, PhysRevB.77.195428, nl3017434, ncomms3010, PhysRevLett.117.166801} or through adatom adsorption.~\citep{PhysRevLett.111.236801, science.aad8038}  Alternatively, the non-trivial $\pi$ electron driven magnetism can be induced at the edges in finite graphene nanoflake (GNF) due to the localized edge states.~\citep{PhysRevB.54.17954, JPSJ.65.1920, Klusek2000508, PhysRevB.71.193406, PhysRevB.73.125415} While the $\pi$ electron is localized at the carbon atoms on zigzag edge, no such localization is expected for armchair edge type.  
While graphene nanostructures have been fabricated,~\citep{nature06037, PhysRevLett.98.206805, Ponomarenko356} the precise control over the 
edge type was limited until few years back. Only very recently using nanofabrication technique and scanning tunneling microscopy, nanoribbons 
with precise zigzag edge have been fabricated, and a robust long-range magnetic order has been observed at room temperature.~\citep{nature13831} Further, semiconductor to metal transition accompanied by switching in the magnetic ordering between the edges has been 
predicted as a function of ribbon width.~\citep{nature13831}  

Graphene is a bipartite hexagonal lattice, which is formed by two interpenetrating triangular sublattices A and B. According to single-band Hubbard model, spins localized at the zigzag edge align in parallel if they belong to the same sublattice. In contrast, they become antiparallel if they belong to different sublattices.~\citep{PhysRevLett.99.177204} Thus, a long-range magnetic ordering is developed along the edge in finite graphene nanoflakes, and the total spin of the ground state follows Lieb's theorem, $2S = N_{\rm A} - N_{\rm B}$.~\citep{PhysRevLett.62.1201}  This picture has been confirmed through density functional theory based calculations in passivated flakes.~\citep{PhysRevLett.99.177204,PhysRevB.90.035403} Further, a magnetic phase transition has been predicted in hexagonal flakes due to carrier doping.~\citep{PhysRevB.90.035403} As the intrinsic magnetic ordering and corresponding ground state magnetization are related to sublattice identity of the edge atoms and total sublattice 
imbalance, an unified size and shape dependence of magnetism in nanoflakes is complex, and not yet fully explored.   

Here, we systematically address this intricate issue within first-principles calculations by considering graphene nanoflakes with 
various shapes and varied sizes. We have considered four shapes, rhombohedral, hexagonal, triangular and pentagonal. 
The nanoflakes with arbitrary shapes can be classified by sublattice imbalance; flakes with sublattice imbalance $N_{\rm A} \ne N_{\rm B}$ (such as triangular and pentagonal), and without imbalance $N_{\rm A} = N_{\rm B}$  (rhomohedral and hexagonal). In order to investigate the $\pi$ electrons only, the in-plane dangling bonds are passivated with single hydrogen atom. For the nanoflakes without any sublattice imbalance, the prediction of Lieb's theorem $S=0$ may be satisfied either with a nonmagnetic solution or with a fully compensated intrinsic magnetic ordering. This may lead to a quantum phase transition with flake size.~\citep{PhysRevLett.99.177204} Indeed, for both rhombohedral (R) and hexagonal (H) shapes, we find that carbon atoms at the edge develop local moments beyond a critical size, and a crossover from nonmagnetic to compensated ferrimagnetic phase takes place. We characterize this with the armchair defect density at the edge, which is defined as the ratio of armchair bond (at the junction of A and B sublattices) with the number of zigzag bonds. In contrast, the flakes with sublattice imbalance, triangular (T) and pentagonal (P) flakes, are always found to be magnetic independent of their respective size.   

Further, manipulation of intrinsic magnetic and corresponding electronic properties by external perturbations is key to spintronic applications. Thus, in addition to undoped flakes, we investigate the carrier doped flakes. Doping could be experimentally achieved by applied gate voltage. In this context, the half-filled Hubbard model in the Nagaoka limit,~\citep{PhysRev.147.392} infinite bipartite lattice with essentially infinite Coulomb interaction, predicts a magnetic phase transition at the filling slightly away from the half-filling. However, the present cases of nanoflakes, neither of the Nagaoka limits are satisfied. Nonetheless, we observe the intrinsic magnetic structure is crucially entangled with the doped carrier density, and complex magnetic phase changes are predicted. Together with changes in magnetic phases, a semiconductor to (half) metal 
transition has been observed. With the current state of advancement in controlled experimental techniques, we hope that our theoretical observations will stimulate further experimental activity.

\section{Computational Details}
Calculations were carried out within density functional theory (DFT) in the plane wave basis, as implemented in the Vienna {\it Ab-initio} Simulation Package (VASP).~\citep{PhysRevB.47.558,PhysRevB.54.11169} The projector augmented wave pseudopotential was used with a plane-wave cutoff of 800 eV.~\citep{PhysRevB.50.17953}  For the exchange-correlation functional, we used the Perdew-Burke-Ernzerhof (PBE) form of generalized gradient approximation (GGA).~\citep{PhysRevLett.77.3865} The reciprocal space integrations were carried out at the $\Gamma$ point. Within the periodic set-up of the calculations, flakes were placed in simple cubic supercells such that the periodic images are separated by at least 12 {\AA} of vacuum space, which ensures the interaction between the images negligible.  Nanoflakes were optimized until the forces on each atom became less than 5 $\times$ 10$^{-3}$ eV/\AA. Further, to understand the effect of electron-electron correlation beyond GGA, some of the calculations were repeated considering a Hubbard like on-site Coulomb interaction, $U$ (DFT+$U$), within rotationally invariant approach.~\citep{PhysRevB.57.1505}

\begin{figure}
\begin{center}
\rotatebox{0}{\includegraphics[width=0.4\textwidth]{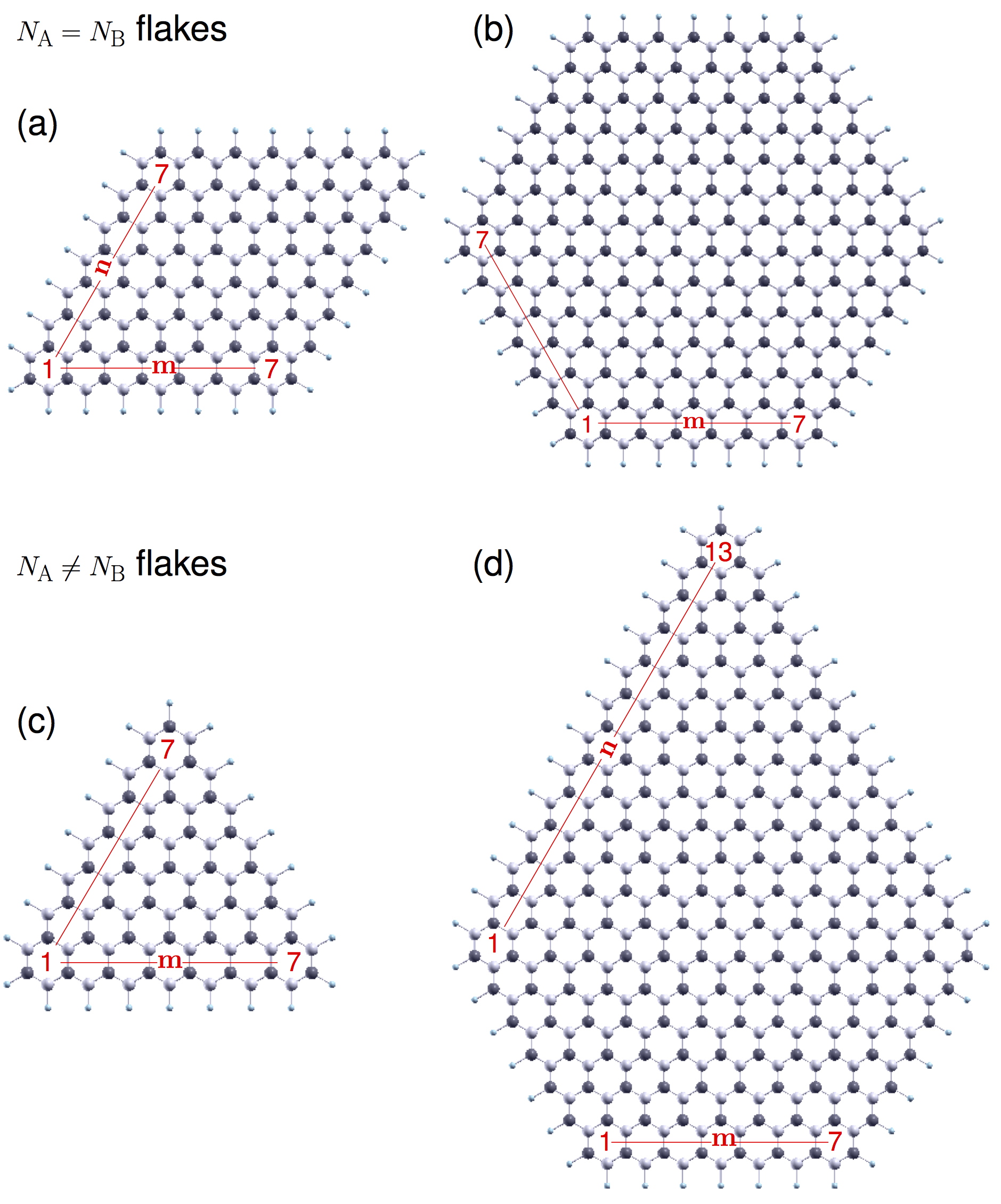}}
\end{center}
\caption{(Color online) Hydrogenated graphene nanoflakes, considered in the present study, (a) Rhombohedral [R], (b) hexagonal [H], (c) triangular [T], and (d) pentagonal [P] flakes. Carbon atoms belonging to two sublattices are marked with black and white colored balls. The H atoms are shown 
with small sized blue balls. The size of the flakes is characterized by the number of hexagonal rings along the edge. While sizes of regular T 
and H flakes are represented by edge lengths $m$, the R and P flakes are represented as $m$$\times$$n$ and $m$$\otimes$$n$, respectively.   
}
\label{figure:1}
\end{figure}

We considered regular T and H flakes with sizes $m$=3, 5, 6, 8; and  3, 6, 7, 8, 9, 10, respectively, where $m$ is the number of hexagonal rings along their regular edge (Figure~\ref{figure:1}). The size of the R and P flakes are denoted by $m\times n$ and $m \otimes n$, where $m$ and $n$ are 
lengths of dissimilar edges (Fig. \ref{figure:1}). While 3$\times$3, 3$\times$5, 3$\times$7, 4$\times$4, 5$\times$5, 7$\times$7 were considered for the R flake, we considered 4$\otimes$7, 5$\otimes$9, 
7$\otimes$13, and 8$\otimes$15 for the P flake.

\begin{figure*}
\begin{center}
\rotatebox{0}{\includegraphics[width=1\textwidth]{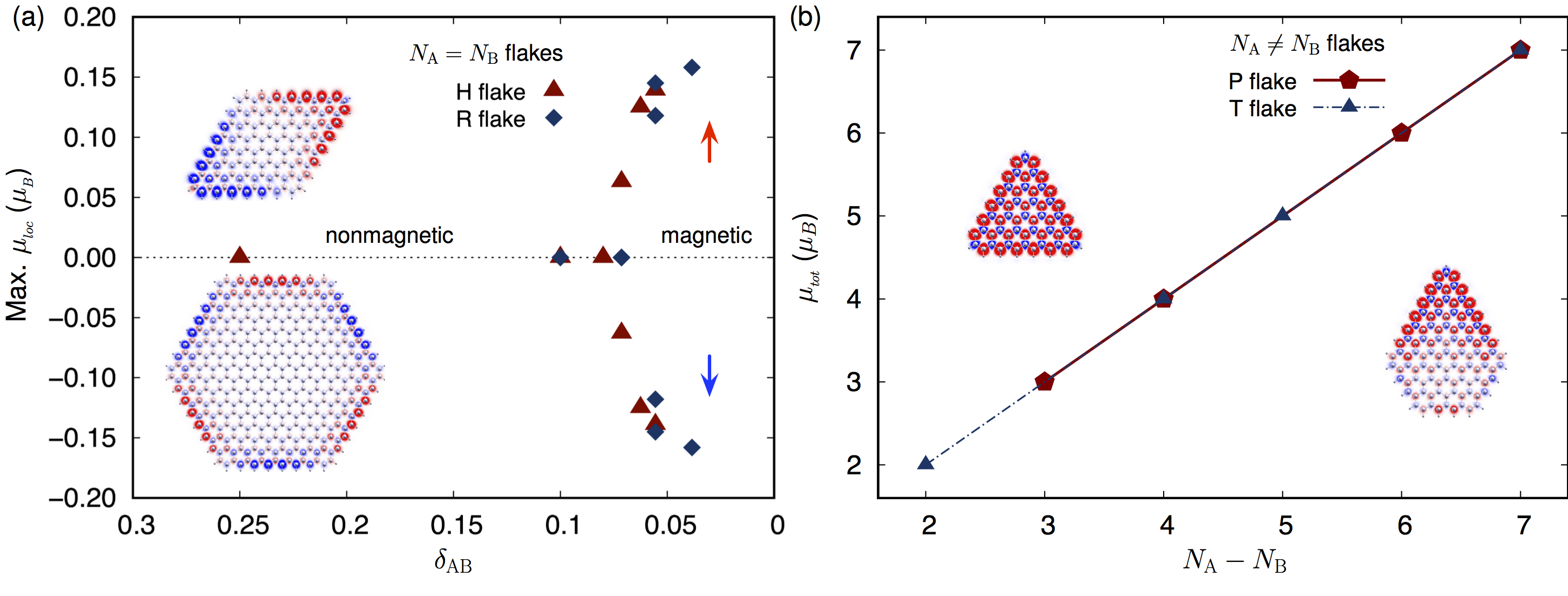}}
\end{center}
\caption{Size and shape dependent magnetism in graphene flakes. (a) For $N_{\rm A}=N_{\rm B}$ flakes, a nonmagnetic to magnetic transition is observed with increasing the flake size characterized by the armchair defect concentration at the edge $\delta_{\rm AB}$.  Variation of calculated maximum local moment $\mu_{loc}$ with  $\delta_{\rm AB}$ indicate a magnetic phase transition at a critical flake size $\delta_{\rm AB}$ $\sim$ 0.07.  (b) In contrast, the $N_{\rm A} \ne N_{\rm B}$ are always magnetic, and total moment of the flake $\mu_{tot}$ follow Lieb's theorem. Representative ground state magnetization densities are shown as insets, where majority (minority) spin densities are shown in red (blue) color.  
}
\label{figure:2}
\end{figure*}

\section{Results and discussion}

We start our discussion with undoped GNFs to illustrate the intricate shape and size dependence of the intrinsic edge magnetism. We also investigate the effect of strong electron correlation on the magnetism. Having understood the intricate size, shape and correlation dependent magnetism in undoped GNFs, we discuss the effect of carrier doping on the magnetic and electronic structures. 
 
\subsection{Shape and size dependence of edge magnetism}

\begin{figure}[b]
\begin{center}
\rotatebox{0}{\includegraphics[width=0.5\textwidth]{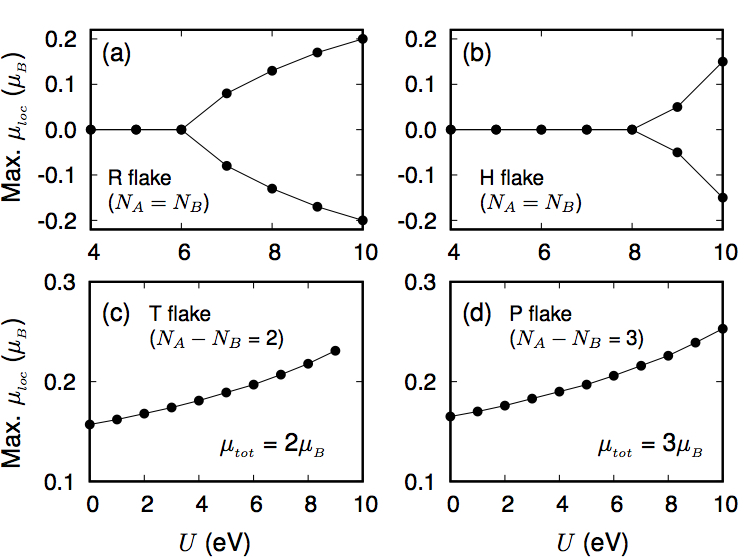}}
\end{center}
\caption{Effect of on-site Coulomb interaction $U$ on the edge magnetism. The maximum $\mu_{loc}$ calculated within DFT+$U$ formalism are shown with varied $U$ for (a) 3$\times$3 R-flake, (b) $m$=3 H flake, (c) $m$=3 T flake, and (d) 4$\otimes$7 P flake.  
}
\label{figure:3}
\end{figure}

We find that the intrinsic magnetic structures of the GNFs are strongly dependent on both nanoflake shape and their size.   
The scenario for the flakes without any sublattice imbalance is particularly non-trivial, and a quantum phase transition is 
observed [Figure~\ref{figure:2}(a)]. For both R and H flakes, we find that only beyond a critical size, the carbon atoms at 
the edge develop a local moment, and a nonmagnetic to fully compensated ferrimagnetic phase change occurs at this critical size. For the magnetic R and H flakes, the intrinsic intra-edge coupling is found to be ferromagnetic as the atoms on a specific edge 
belong to particular sublattice. In contrast, antiferromagnetic coupling is observed between the edges, which belong to different sublattices [Figure~\ref{figure:2}(a)]. This picture is consistent with the predictions from the Hubbard model, and agrees well with the previous DFT calculations for H flakes.~\citep{PhysRevLett.99.177204,PhysRevB.90.035403} However, the origin of this size dependence has not been addressed previously. To characterize the observed magnetic phase transition with flake size, we define a unitless quantity $\delta_{\rm AB}$, which relates to the armchair defects along the entire edge of the flake. There exists an armchair bond at the junctions of A and B sublattice [Figure~\ref{figure:1} (a)], where electrons are delocalized, and thus is detrimental for magnetism. Thus, in the context of magnetism on zigzag edge, such armchair bonds act as defects. This is also evident from the magnetization density. The local moment $\mu_{loc}$ along a particular edge has a distribution -- zero at the armchair defect, and increases away from the armchair defects at the junctions of A and B sublattices [Figure~\ref{figure:2}(a)]. Thus, for H flakes, the carbon atoms at the middle of the edge has the maximum local moment. In comparison, for R flakes, $\mu_{loc}$ progressively increases in moving towards the defect free corners, starting from the zero value at two armchair defects connecting A and B sublattices. We define the defect density as $\delta_{\rm AB} = N_{\rm arm}/N_{\rm tot}$, where $N_{\rm arm}$ ($N_{\rm tot}$) is the number of armchair (total) bonds along the edge. Thus,  $\delta_{\rm AB}$ decreases with increasing flake size. We find that beyond a critical flake size with corresponding $\delta_{AB} < $ 0.07, the edge atoms develop a moment, and the flakes become compensated ferrimagnet [Figure~\ref{figure:2}(a)].  It is important to note that with increasing flake size ($\delta_{AB} < $ 0.07) the maximum local moment at the edge, $\mu_{loc}$ increases, but keeping the net moment fixed at zero value, thus obeying the prediction of Lieb's theorem.

In comparison, the size dependence of magnetism in T and P flakes having sublattice imbalance is far less complex. Independent of size, such $N_{\rm A} \ne N_{\rm B}$ flakes are found to be magnetic with the total ground state  magnetization given by, $\mu_{tot}=2S=N_{\rm A} - N_{\rm B}$, following Lieb's theorem [Figure~\ref{figure:2}(b)]. Thus, with increase in size, the sublattice imbalance increases, leading to linear increase in $\mu_{tot}$. For T flakes, all the edge atoms belong to a particular sublattice, and thus the inter-edge coupling is ferromagnetic [Figure~\ref{figure:2}(b)]. The P flakes can be viewed as a fusion of T and H structures. While the edge atoms in the triangular region belong to the same sublattice, rest of the edges alternate between two sublattices [Figure~\ref{figure:1}(c) and  Figure~\ref{figure:1}(d)]. The corresponding magnetic structure is thus commensurate with the ferromagnetic and antiferromagnetic inter-edge coupling between the same and different sublattices [insets in Figure~\ref{figure:2}(b)].


Next, we study the effect of strong electron correlation on the edge magnetism, as a strong Coulomb interaction has been argued for graphene nanostructures.~\citep{PhysRevLett.106.236805,nature08942,0295-5075-19-8-007,PhysRevB.72.085123} The local Coulomb interaction is proposed to be as large as $\sim$ 9 eV, which is in between the values that predicts a spin-liquid phase~\citep{nature08942} and a compensated antiferromagnetic phase.~\citep{0295-5075-19-8-007,PhysRevB.72.085123} Keeping this in mind, we have repeated our calculation by supplementing a Hubbard like on-site Coulomb interaction $U$ (0-10 eV) on the conventional PBE exchange-correlation. Only one representative size has been considered for each shape (Figure~\ref{figure:3}).

For the flakes without any lattice imbalance in the size range $\delta_{\rm AB} > $ 0.07, the flakes are nonmagnetic, and with increase in on-site Coulomb interaction beyond a critical value the edge atoms develop a local moment [Figure~\ref{figure:3} (a) and (b)]. Further, $\mu_{loc}$ increases with $U$, while the complete magnetic structure remains fully compensated ferrimagnet. This picture is similar to that with increasing flake size, which has been discussed in the previous section. Thus, we conclude that increasing size and increasing electron-electron correlation both help in enhancing the localized character of the edge states, thereby stabilizing magnetism. In contrast, in the presence of lattice imbalance, the flakes are always magnetic, and the increase in $U$ monotonically increases the local moment at the edge. However, the ground state magnetization is maintained at $\mu_{tot} = N_{\rm A} - N_{\rm B}$ by compensated increase in A and B sublattice moments. This in turn confirms the validity of Lieb's theorem irrespective of the level of theoretical approximation toward electron correlation. Due the absence of accurate estimate of $U$ for finite sized nanoflakes, and due to the fact that the flake size and $U$ affect the intrinsic magnetic structure in similar fashion, we ignore explicit incorporation of Coulomb interaction for rest of the paper. Rather, we will consider larger flakes with $\delta_{\rm AB} < $ 0.07, where are flakes are already magnetic without explicit incorporation of electron correlation.

\begin{figure}
\begin{center}
\rotatebox{0}{\includegraphics[width=0.4\textwidth]{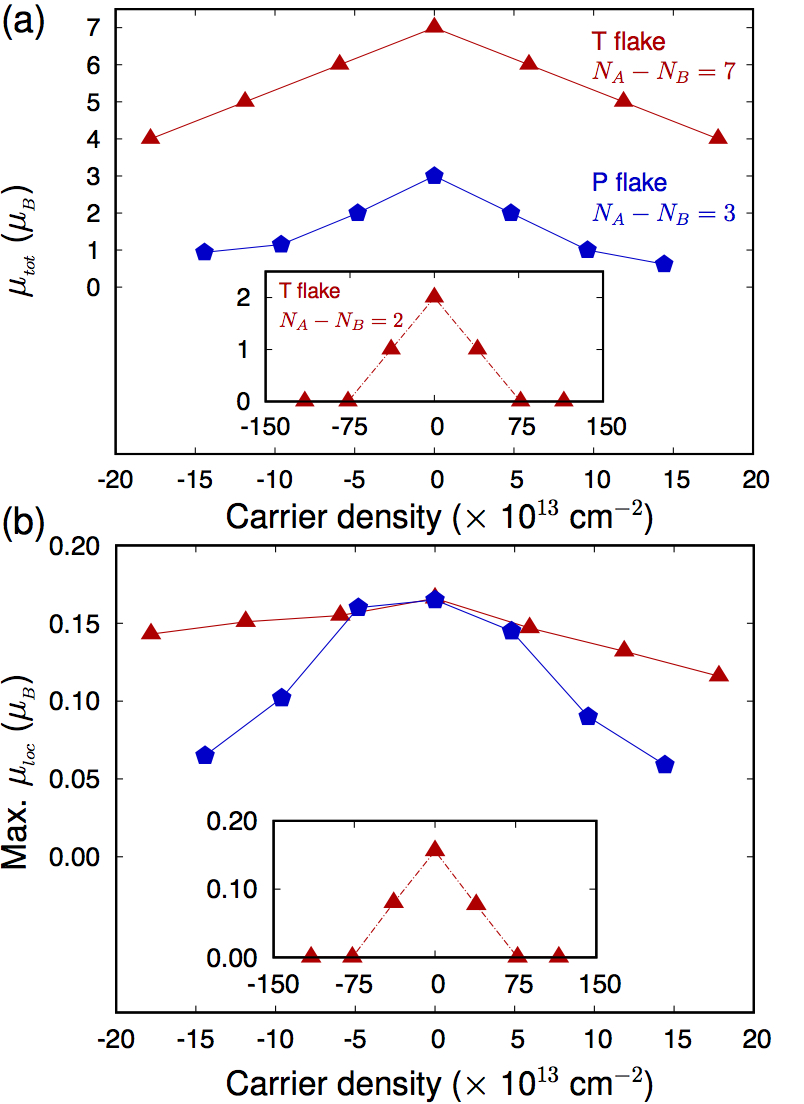}}
\end{center}
\caption{Effect of carrier (electron and hole) doping on magnetism for $N_{\rm A} \ne N_{\rm B}$ flakes. Calculated (a) total moment $\mu_{tot}$, and (b) maximum local moment on the edge $\mu_{loc}$ with varied carrier doping for T flakes with $m$ = 3 and 8; and 4$\otimes$7 P flake. Positive (negative) doping indicates electron (hole) doping.}
\label{figure:4}
\end{figure}

\subsection{Effects of carried doping on edge magnetism}

Next, we investigate the effect of carrier doping on the edge magnetism for nanoflakes with different shapes. The doped carrier density can be tuned experimentally, and possible manipulation of magnetic structure through gate voltage is fundamental to spintronic applications. We have considered both electron and hole doping with concentrations $\rho_c$ in the range 10$^{13}$ $-$ 10$^{14}$ cm$^{-2}$.  Such high $\rho_c$ has been experimentally achieved.~\cite{ncomms3010, 1.3483130, Ye09082011} Similar to the undoped flakes, effect of carrier doping is found to be strongly shape dependent with sharp distinction in behavior of flakes with and without sublattice imbalance. Additionally, a magnetic phase transition is observed due to carrier doping, which is fundamentally interesting and could be of technological importance.   

We first discuss the flakes with $N_{\rm A} \ne N_{\rm B}$, which in undoped situation is always magnetic, independent of size [Figure~\ref{figure:2}(b)]. As discussed, the inter-edge coupling in T-flakes is ferromagnetic with ground state magnetization $\mu_{tot} = N_{\rm A} - N_{\rm B}$. Thus, the undoped T flakes with $m$= 3, and 8 have 2 and 7 $\mu_B$ moment [Figure~\ref{figure:4}(a)]. Electron (hole) doping in these flakes decreases the value of $\mu_{tot}$ to $N_{\rm A} - N_{\rm B} - n^{e(h)}$, where $n^{e(h)}$ is the number of doped electron (hole). Thus, $\mu_{tot}$ monotonically decreases with increasing carrier doping [Figure~\ref{figure:4}(a)], and $m=3$ flake becomes nonmagnetic beyond a critical $\rho_c$. For larger $m$=8 T flake, $\mu_{tot}$ decreases to 4 $\mu_B$ due to 3 electron (hole) doping, which correspond a carrier doping of 1.79 $\times$ 10$^{14}$ cm$^{-2}$. The P flake magnetism upon carrier doping is found to be qualitatively similar to that of the T flakes. 

\begin{figure}[t]
\begin{center}
\rotatebox{0}{\includegraphics[width=0.4\textwidth]{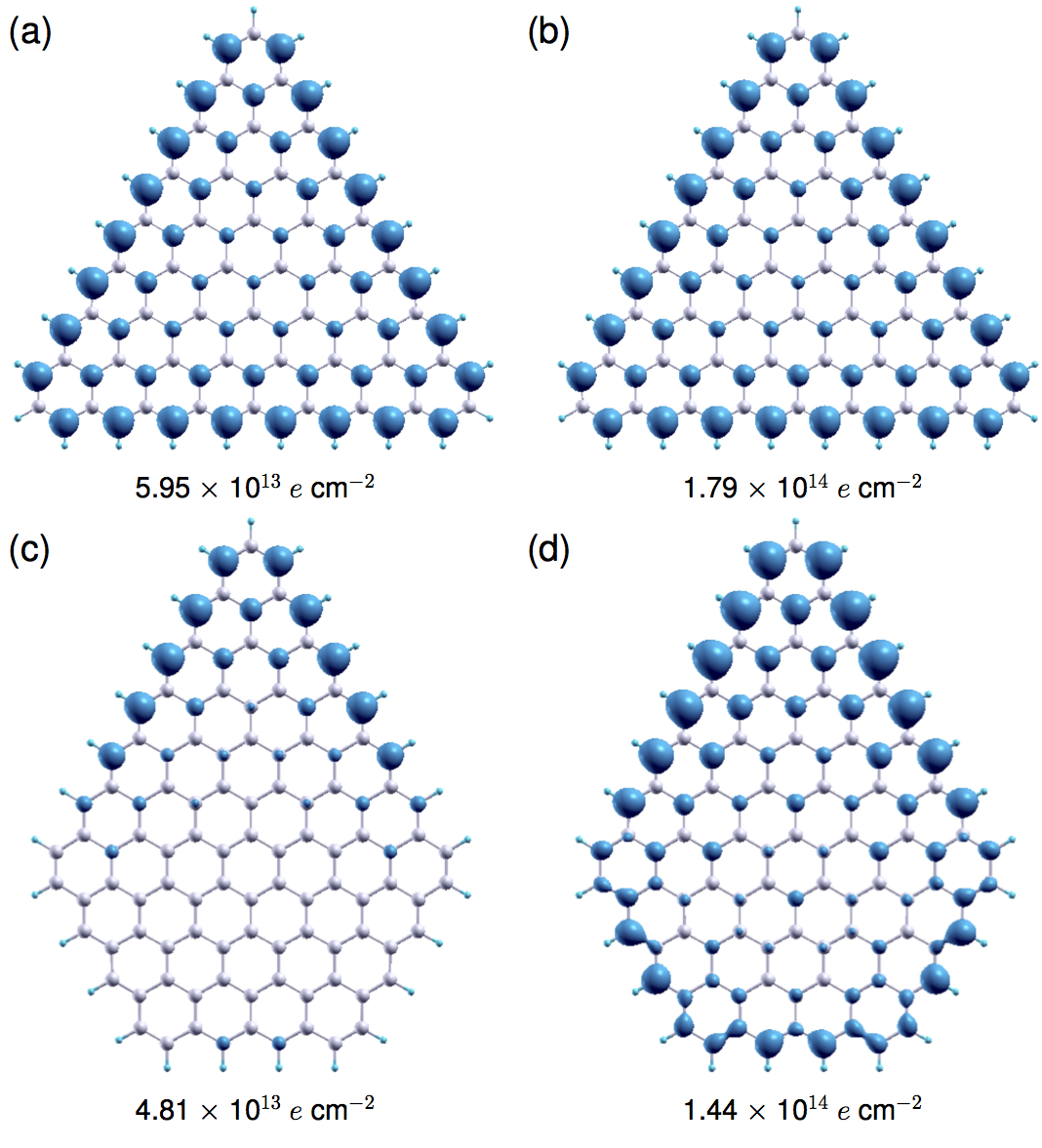}}
\end{center}
\caption{Partial charge density distribution for doped carrier (electron doping) for T flake for two different
carrier densities [(a) and (b)], and for P flake for two different carrier densities [(c) and (d)].}
\label{figure:New}
\end{figure}

To understand the microscopic origin of such behavior in the flakes, we calculated $\mu_{loc}$ for the edge atoms [Figure~\ref{figure:4}(b)], and investigated the partial charge density of the doped electron (hole), shown for $m$=8 T flake and 4$\otimes$7 P flakes [Figure~\ref{figure:New}(a) -- (d)]. We find that doped carrier is distributed mainly on the edge atoms, and entirely on A sublattice. Further, they occupy the minority $p_z$ channel, which leads to a decrease in calculated $\mu_{tot}$ with carrier doping [Figure~\ref{figure:4}(b)]. Concurrently, $\mu_{tot}$ decreases by $n^{e(h)}$ $\mu_B$ from the neutral ground state. However,  such a picture breaks down at higher doping for the P flakes, and the corresponding ground state spin is higher than expected $N_{\rm A} - N_{\rm B} - n^{e(h)}$ as discussed before. At higher doping limit beyond 9.62 $\times$ 10$^{13}$ cm$^{-2}$,  although the carrier is predominantly distributed over the A sublattice, a fraction populates the minority $p_z$ channel of the B sublattice. Thus, the decrease in $\mu_{tot}$ becomes slower than expected [Figure~\ref{figure:4}(a)].  Interestingly for all $N_{\rm A} \ne N_{\rm B}$ flakes, though the calculated $\mu_{loc}$ and $\mu_{tot}$ decreases upon carrier doping, the intrinsic intra-edge and inter-edge magnetic coupling remain same as for the undoped flakes.

 \begin{figure}[t]
\begin{center}
\rotatebox{0}{\includegraphics[width=0.45\textwidth]{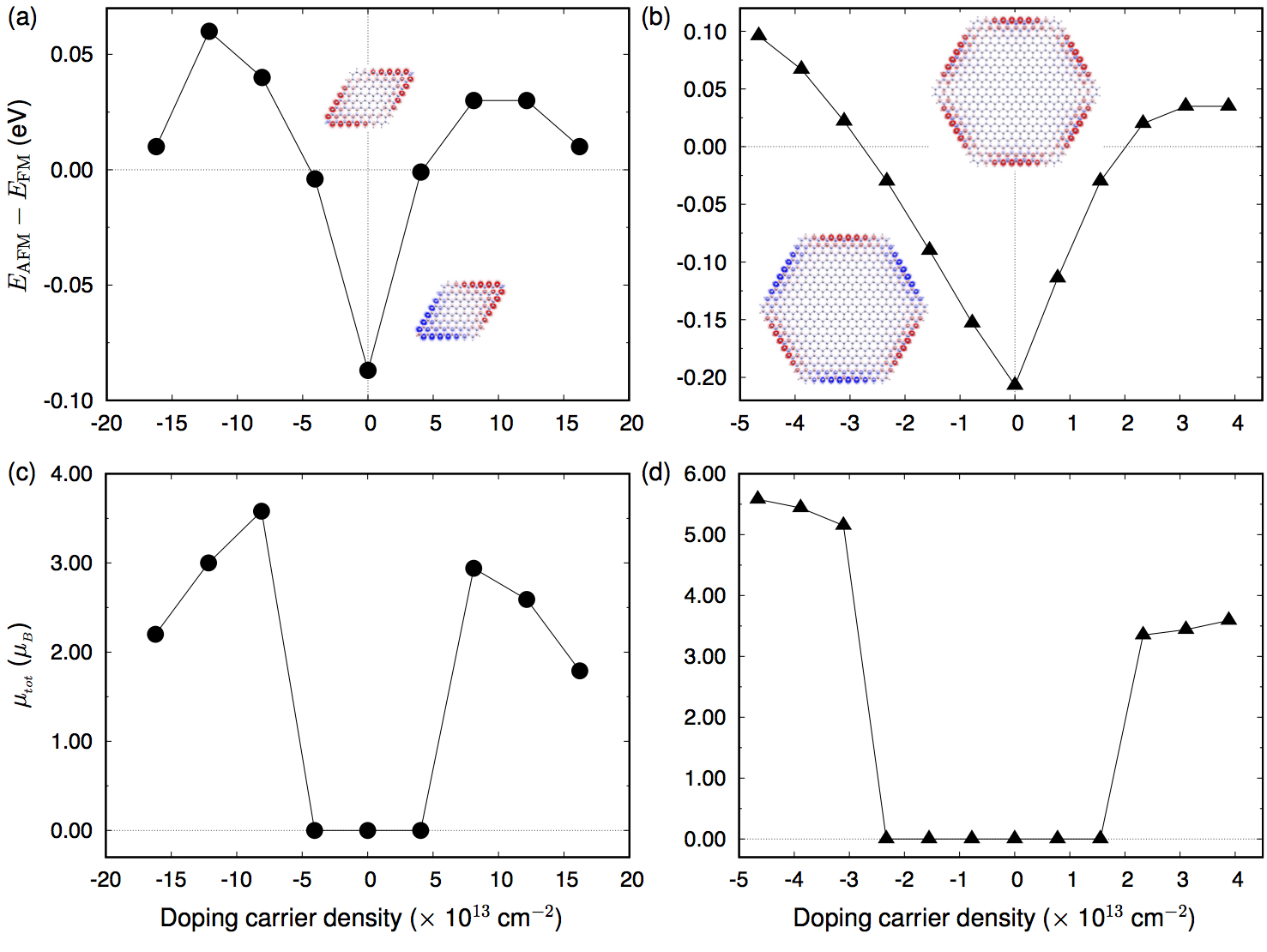}}
\end{center}
\caption{Effect of carrier doping on magnetic $N_{\rm A} = N_{\rm B}$ flakes. Calculated energy differences between the compensated ferrimagnetic (AFM) and ferromagnetic (FM) solutions for (a) 7$\times$7 R flake, and (b) $m$=10 H flake. The insets show representative magnetization densities. The corresponding total moments $\mu_{tot}$ are shown in (c) and (d).}
\label{figure:5}
\end{figure}
 
In contrast, carrier doping affects the magnetism in $N_{\rm A}=N_{\rm B}$ flakes in a non-trivial fashion.  As we have discussed earlier that a nonmagnetic to compensated ferrimagnetic phase transition occurs at a critical size characterized by $\delta_{\rm AB} <$ 0.07 [Figure~\ref{figure:2}(a)].  Here, we consider R and H flakes with $\delta_{\rm AB} <$ 0.07, which is magnetic in their neutral state, and investigate the effect of carrier doping. Interestingly, we find a magnetic phase transition upon carrier doping beyond a critical $\rho_c$ [Figure~\ref{figure:5}(a) and (b)]. The inter-edge coupling becomes ferromagnetic with non-zero $\mu_{tot}$ [Figure~\ref{figure:5}(c) and (d)]. Thus, the impact of carrier doping is substantially different in these flakes than for the flakes with sublattice imbalance. While, carrier doping affects $\mu_{loc}$, and thus $\mu_{tot}$ in  $N_{\rm A} \ne N_{\rm B}$ flakes without affecting the nature of long-range order, a phase transition is observed for $N_{\rm A}=N_{\rm B}$ flakes. Further, electron-hole asymmetry is observed in terms of stabilization of a particular magnetic phase [Figure~\ref{figure:5} (a) and (b)], although the phase transition is induced by  both electron and hole doping. In case of R flakes, we notice a plausible re-entrance to compensated ferrimagnetic phase as the energy difference $E_{\rm AFM}-E_{\rm FM}$ decreases at large $\rho_c$.

\subsection{Effects of carried doping on electronic structure}

\begin{figure}
\begin{center}
\rotatebox{0}{\includegraphics[width=0.45\textwidth]{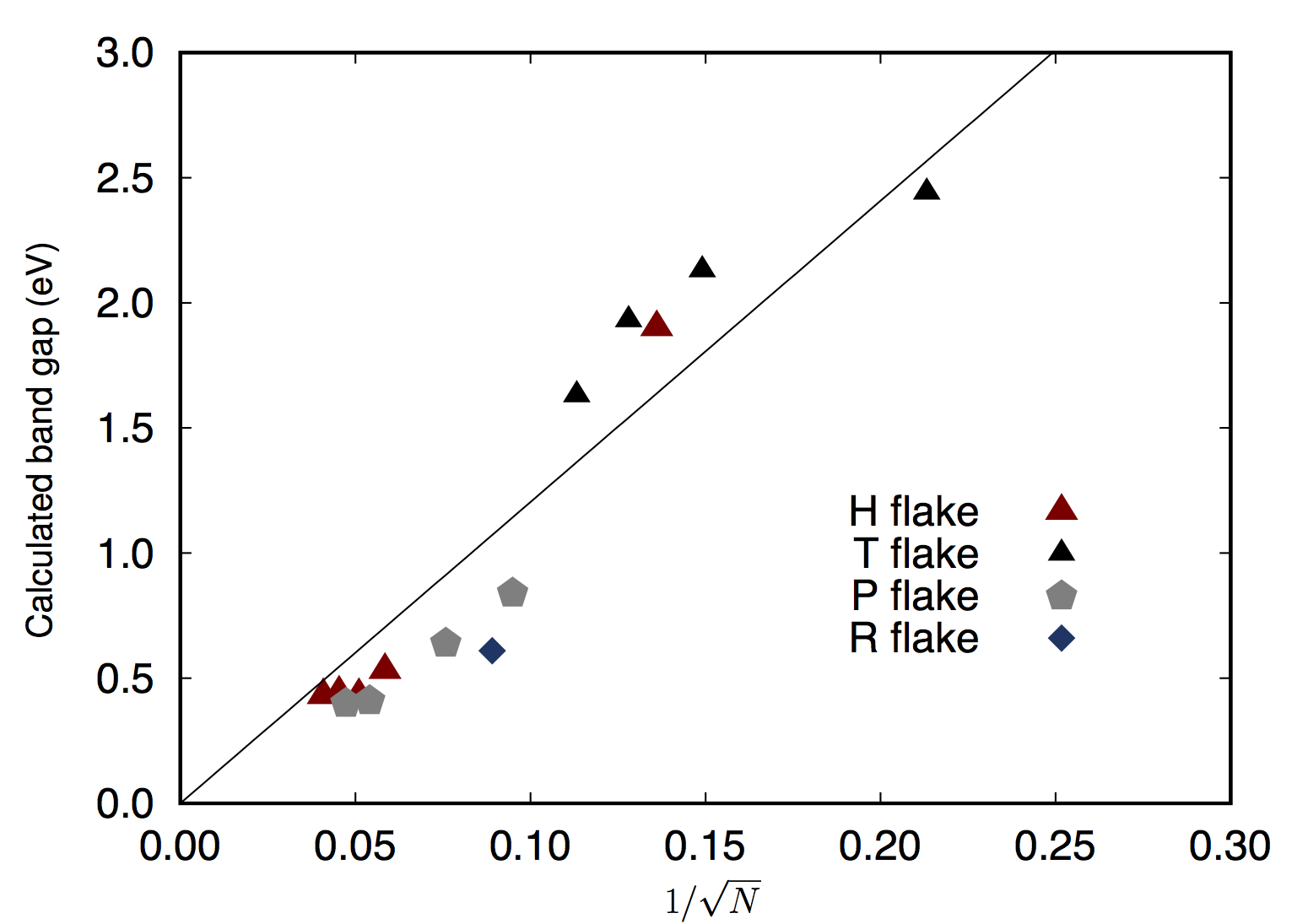}}
\end{center}
\caption{Calculated band gap for the flakes with different shape and size, where $N$ is the number of carbon atoms in the respective nanoflake. The solid line is a linear fit indicating that the band gap decreases with increasing flake size. 
}
\label{figure:gap}
\end{figure}

Although bulk graphene is a semimetal with zero density of states at the Dirac points, the electronic structure of graphene quantum dots and ribbons depend sensitively on the crystallographic orientation of their edges.~\citep{nature05180,PhysRevLett.98.206805,Ponomarenko356, nl0617033,PhysRevLett.97.216803,Li1229,jcp1.4865414,jcp1.4902806} The neutral flakes are found to be semiconducting and the calculated gaps decrease with increasing flake size, which we characterize by the total number of atoms $N$ in the flake (Figure~\ref{figure:gap}). Although, the gap should go to zero at infinite size, the asymptotic interpolation might be complex,~\citep{jcp1.4865414,jcp1.4902806} due to intricate dependence on the flake shape, and corresponding magnetism. However, regardless the shape, the neutral flakes are found to be semiconducting in the studied size range (Figure~\ref{figure:gap}). Thus, it would be interesting to investigate the effect of carrier doping on the semiconducting nature of these flakes.  

\begin{figure}[t]
\begin{center}
\rotatebox{0}{\includegraphics[width=0.4\textwidth]{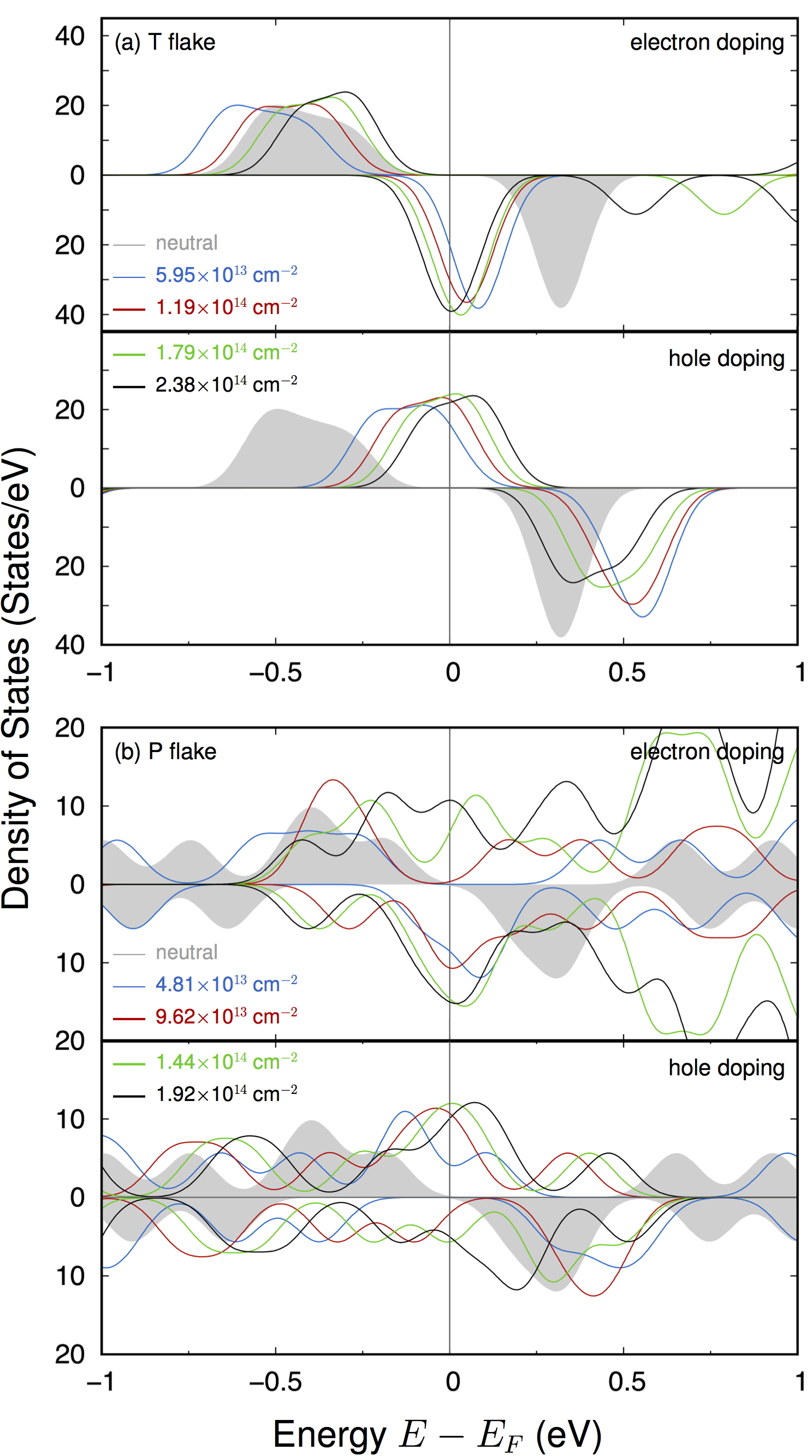}}
\end{center}
\caption{Evolution of the electronic structure with carrier doping for (a) $m$=8 T flake, and (b) 4$\otimes$7 P flake. While a semiconductor - half-metal transition is observed for the T-flake, and semiconductor - half-metal - metal transition is observed for the P flake.}
\label{figure:ES1}
\end{figure}

We observe an electronic phase transition in these flakes, and semiconducor to (half) metal transition is observed due to carrier doping. We first investigate the $N_{\rm A}=N_{\rm B}$ flakes in detail, where we particularly consider the $m$=8 T flake and 4$\otimes$7 P flakes.  The neutral $m$=8 T flake is a semiconductor with 1.29 (1.27) eV gap in the majority (minority) channel. Careful investigation of the density of states (DOS) indicate that the highest occupied majority $p_z$ states lie 0.10 eV below the Fermi level $E_F$, whereas the unoccupied minority $p_z$ states lie 0.11 eV above $E_F$. Single electron (hole) doping with corresponding $\rho_c$=5.95$\times$10$^{13}$ cm$^{-2}$ alters the electronic structure and transform the flake into half-metallic [Figure~\ref{figure:ES1} (a)]. The flake remains half-metallic upon further doping. 

The otherwise unoccupied minority $p_z$ channel gets populated by electron doping [Figure~\ref{figure:ES1} (a)], and thus the minority channel becomes conducting. In contrast, for hole doping, electrons are removed from the majority $p_z$ channel, and thus $E_F$ moves lower in energy. This leads to a gap closing in the majority channel [Figure~\ref{figure:ES1} (a)]. While increase in electron doping monotonically increases the DOS at $E_F$, for hole doping the $E_F$ is continuously pushed toward lower energy. Therefore, the overall qualitative picture remains similar with the increase in $\rho_c$, and the solution remains half-metallic [Figure~\ref{figure:ES1} (a)] at all $\rho_c$.  Thus, T flakes will manifest completely spin-polarized transport under applied gate voltage. Further, it would be possible to control conduction in a particular channel (majority or minority) by controlling the polarity of the gate voltage (electron or hole doping). This would be significant importance in the context
of spin-based electronics.

\begin{figure}[t]
\begin{center}
\rotatebox{0}{\includegraphics[width=0.4\textwidth]{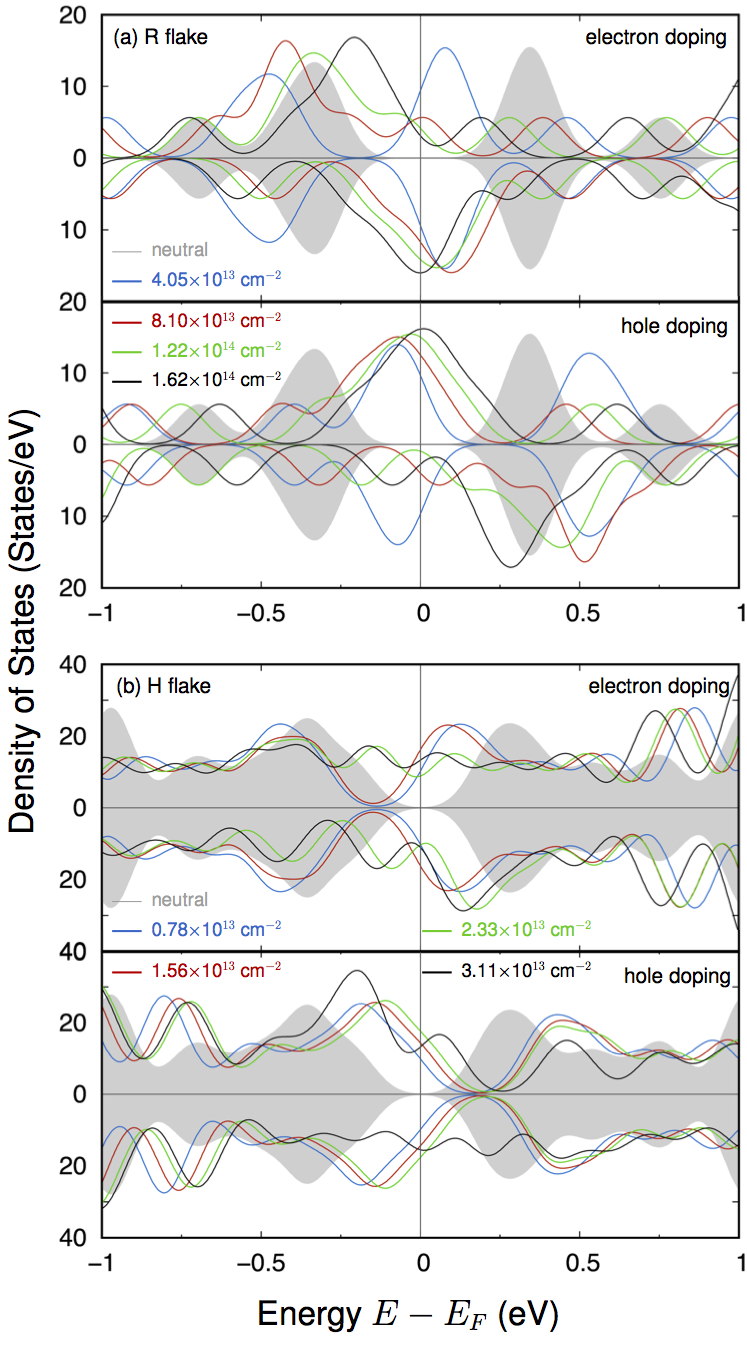}}
\end{center}
\caption{Carrier induced electronic structure evolution for (a) 7$\times$7 R flake, and (b) $m$=10 H flake. For both the flakes a 
semiconductor - metal transition is observed due to electron/hole doping.}
\label{figure:ES2}
\end{figure}

The carrier induced evolution of electronic structure in P flakes is qualitatively different. A semiconductor to half-metal to metal transition is observed with increasing carrier density. The neutral 4$\otimes$7 P flake is semiconducting with 0.55 (0.51) eV gap in the majority (minority) channel. While the highest occupied $p_z$ states appear 0.05 eV below the $E_F$, the lowest unoccupied $p_z$ states 0.05 above the $E_F$. For electron doping with $\rho_c \leqslant$ 9.62$ \times$ 10$^{13}$ cm$^{-2}$, the doped electron populates the minority channel only, and the flake becomes half-metallic [Figure~\ref{figure:ES1} (b)]. Further increase in electron density populates both majority and minority channels simultaneously, and a metallic solution emerges for $\rho_c \geqslant$ 1.44$\times$ 10$^{14}$ cm$^{-2}$. In contrast, for hole doping, electrons are removed from the majority channel only for $\rho_c \leqslant$ 9.62$\times$ 10$^{13}$ cm$^{-2}$.  As the $E_F$ moves lower in energy, a half-metallic solution emerges. Further increase in hole doping depletes electrons from both the channels, and the ground state becomes metallic. This picture also explains the emergence of non-integer $\mu_{tot}$ for the corresponding ground state, and also illustrates the slow decrease in $\mu_{tot}$ for the metallic P flakes, at high carrier doping. 

We next discuss the carrier dependent electronic structure for the flakes without sublattice imbalance, $N_{\rm A} = N_{\rm B}$. Here, we consider a 7$\times$7 R flake, and a $m$=10 H flake, which show fully compensated ferrimagnetic structure. These flakes have 240 and 200 meV semiconducting gaps, respectively, in their neutral states. Independent of flake shape, a semiconductor to metal transition is observed [Figure~\ref{figure:ES2}(a) and (b)], without any appearance of half-metallic solution. As the inter-edge coupling is antiferromagnetic in these flakes, the doped electron equally populates the majority and minority channels. Thus, gaps in both the channels disappear and the flakes become metallic. Similarly, the electrons are depleted from both the channels due to hole doping. Thus the $E_F$ continuously shifts to a lower energy and the solutions become metallic. Similar semiconductor to metal transition has been experimentally observed in graphene nanoribbons.~\citep{nature13831}

\section{Summary}
Using first-principles calculations, we carried out a systematic study of the magnetic and electronic structures of graphene nanoflakes with different shapes and sizes. Further, the effect of carrier doping was investigated, which could be accessed experimentally by gate voltage.  The presence or absence of sublattice imbalance plays a crucial role in magnetism. For flakes without any sublattice imbalance, a nonmagnetic to magnetic phase transition is observed with increasing flake size, which is characterized by the armchair defect concentration $\delta_{\rm AB}$ along the edge. We find that  beyond a critical flake size $\delta_{\rm AB} <$ 0.13, the carbon atoms at the edge develop local moments and the fake becomes magnetic. In contrast, the flakes with sublattice imbalance are always found to be magnetic. Further, for the neutral flakes with any shape, the edge moments from the same sublattice couple ferromagnetically, while antiferromagnetic coupling is observed between the moments from a different sublattice. This is in agreement with the mean-field solution of the Hubbard model, and in all cases the ground state magnetization follow Lieb's prediction, $\mu_{tot}=(N_{\rm A}-N_{\rm B})$. 

We find that the magnetic phase transition is induced by carrier doping. For $N_{\rm A}=N_{\rm B}$ flakes, beyond a critical doping antiferromagnetic to ferromagnetic phase transition is observed. In contrast, carrier doping suppresses the magnetism in $N_{\rm A} \ne N_{\rm B}$ flakes. In these flakes, the local moments at the edge, and thus $\mu_{tot}$ monotonically decreases with increasing doping. This indicates that at very high doping density the flakes may become nonmagnetic. Further, a simultaneous electronic phase transition is observed in response to carrier doping. In this regard, the cases with sublattice imbalance is found to be more interesting. A semiconductor to half-metal, and a semiconductor to half-metal to metal transition is observed in T and P flakes, respectively. In particular, the half-metallic solution is interesting, and may lead to fully polarized transport. In contrast, for the $N_{\rm A}=N_{\rm B}$ flakes, a semiconductor to metal transition take place without an appearance of half-metallic solution. Finally, we propose that nano-lithographic technique coupled with scanning tunneling microscopy will be able to 
verify our theoretical prediction of complex size and shape-dependent magnetic and electronic phase transitions in graphene flakes, which 
may lead to electronic and spintronic applications.

\begin{acknowledgements}
Some of the reported calculations have been done using the supercomputing facilities at the Centre for Development of Advanced Computing, Pune; Inter University Accelerator Centre, Delhi; and at the Center for Computational Materials Science, Institute of Materials Research, Tohoku University. M. K. acknowledges the funding from the Department of Science and Technology, Government of India under Ramanujan Fellowship, and Nano Mission project SR/NM/TP-13/2016(G).
\end{acknowledgements}

%

\end{document}